\documentclass{optica-article}

\journal{opticajournal} 

\articletype{Research Article}

\usepackage{lineno, soul, color, comment, amsmath}

\usepackage{float}
\usepackage{amsmath}
\usepackage{amsfonts}

\begin{document}

\title{Dynamic Spectral fluorescence microscopy via Event-based \& CMOS image-sensor fusion}

\author{Richard G. Baird,\authormark{1} Apratim Majumder,\authormark{1} and Rajesh Menon\authormark{1,*}}

\address{\authormark{1} Department of Electrical \& Computer Engineering, University of Utah, 50 Central Campus Dr., Salt Lake City, UT 84112, USA\\}

\email{\authormark{*}rmenon@eng.utah.edu} 


\begin{abstract*} 
We present a widefield fluorescence microscope that integrates an event-based image sensor (EBIS) with a CMOS image sensor (CIS) for ultra-fast microscopy with spectral distinction capabilities. The EBIS achieves temporal resolution of $\sim10\thinspace\mu$s ($\sim\thinspace$50,000 frames/s), while the CIS provides diffraction-limited spatial resolution. A diffractive optical element encodes spectral information into a diffractogram, which is recorded by the CIS. The diffractogram is processed using a deep neural network to resolve the fluorescence of two beads, whose emission peaks are separated by only 7 nm and exhibit an 88\% spectral overlap. We validate our microscope by imaging the capillary flow of fluorescent beads, demonstrating a significant advancement in ultra-fast spectral microscopy. This technique holds broad potential for elucidating foundational dynamic biological processes.
\end{abstract*}

\section{Introduction}
Fluorescence microscopy has been instrumental in advancing our understanding of cellular and molecular dynamics. Achieving high temporal resolution in fluorescence microscopy is crucial for capturing fast biological processes, yet it presents significant technical challenges. One of the primary obstacles is the need for sensors that can capture rapid changes in fluorescence intensity with minimal latency, which demands advancements in sensor technology, such as event-based sensors and high-speed CMOS image sensors. Additionally, managing the trade-off between temporal resolution and signal-to-noise ratio is crucial, as higher temporal resolutions often result in reduced signal quality. Computational methods, including machine learning algorithms, are essential to reconstruct high-resolution temporal data from noisy, low-resolution measurements. Integrating these sophisticated sensors and computational techniques while maintaining biological sample viability and minimizing phototoxicity further complicates the development of high temporal resolution fluorescence microscopy systems.

High temporal resolution fluorescence microscopy holds significant promise for advancing various biological applications, as varied as study of synaptic vesicle dynamics to tracking plankton motion.\cite{takatsuka2024millisecond} For instance, temporal resolutions of several tens of milliseconds have been achieved in prior studies\cite{miki2020direct}. Many critical biological phenomena, such as the rapid kinetics of vesicle fusion and neurotransmitter release in neurons, occur on much shorter timescales. Therefore, further improvements in temporal resolution are required to fully capture these fast dynamics. Similarly, calcium imaging has been achieved typically for large neuronal populations, but at temporal scales of about 100 ms,\cite{demas2021high} although it is well known that transient calcium fluxes, providing insights into neuronal activity and signal transduction pathways, can occur much faster\cite{bootman2020fundamentals}. Two-photon microscopy of calcium imaging at about 1ms temporal precision was demonstrated with specialized acousto-optic modulators and signal processing\cite{grewe2010high}. In cardiac research, machine learning has facilitated the real-time visualization of myocyte contractions with temporal resolution of several hundred milliseconds,\cite{psaras2021caltrack} improving our understanding of cardiac function and disease. Achieving sub-millisecond temporal resolution with high spatial resolution can significantly enhance our understanding of numerous biological phenomena, including vesicle trafficking, ion-channel activity, and intracellular signaling. Prior work in spectrally-resolved fluorescence imaging has used sophisticated SPAD sensors for imaging dental caries\cite{kekkonen2023time}. However, such approaches cannot be readily scaled to closely overlapping emission spectra and high temporal resolution. 

Recent advancements in event-based imaging have leveraged the capabilities of event-based image sensors (EBIS) to achieve significant breakthroughs in microscopy. For instance, EBIS has been employed in event-based light field microscopy, surpassing the limitations of conventional CMOS systems and enabling ultrafast 3D imaging at kHz frame rates \cite{guo2024eventlfm}. Similar strides have been made using structured illumination for high-speed 3D microscopy \cite{fu2023fast}. EBIS has also been applied to single-molecule localization super-resolution microscopy, capturing blinking events with enhanced efficiency and dynamic range \cite{cabriel2023event,basumatary2023event}. Furthermore, neuromorphic event sensing has been used to achieve millisecond-scale autofocusing by detecting sparse brightness changes and quickly responding to specimen movement \cite{ge2023millisecond}. However, to our knowledge, no prior studies have demonstrated spectrally resolved microscopy on the timescales accessible by EBIS. 

\begin{figure}[htb!]
\centering\includegraphics[width=12cm]{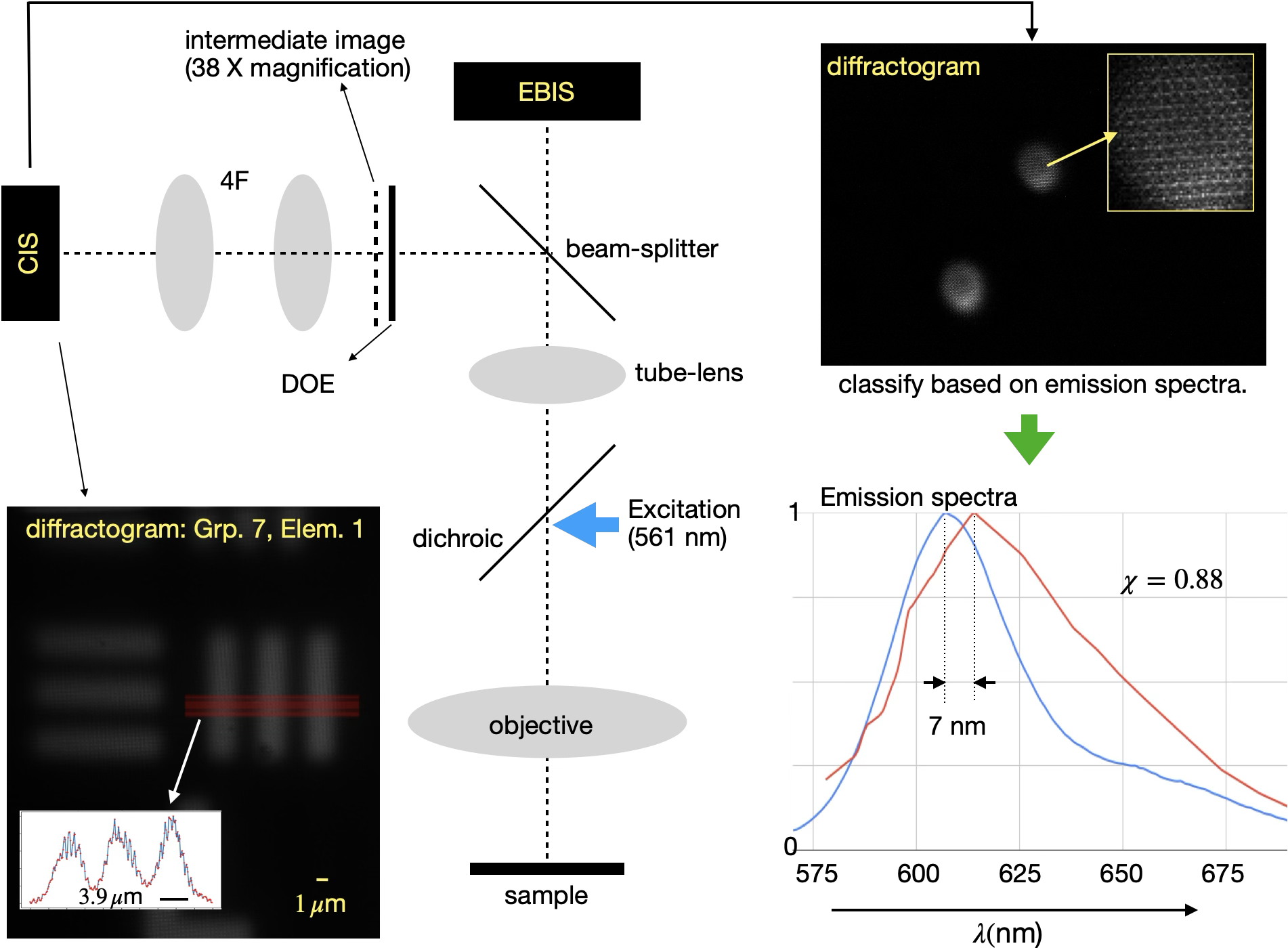}
\caption{Optical setup of the event-based image sensor (EBIS) - CMOS image sensor (CIS) widefield fluorescence microscope. A diffractive optical element (DOE) is positioned near the intermediate image plane, producing a spectrum-sensitive image (diffractogram), which is relayed onto the CIS by a 4f system. This diffractogram (top-right inset shows magnified view) is then analyzed by a pre-trained classifier to identify the beads based on their fluorescence spectra. Note that the fluorescence spectra of the two beads exhibit significant overlap (correlation, $\chi=$ 88\%) and the emission peaks are separated only by 7 nm. Lower-left inset shows the diffractogram of an AirForce resolution chart, group 7, element 1 and the corresponding line-scans indicating good contrast ($\sim 68\%$) for lines of width $3.9\thinspace\mu$m. Effective magnification is $38\times$. Only a portion of the full field-of-view of $54 \times 124\thinspace\mu$m is shown.}
\label{fig:setup}
\end{figure}
In this study, we present the integration of an event-based image sensor (EBIS) with a CMOS image sensor (CIS) in a conventional widefield fluorescence microscope to attain enhanced temporal, spatial, and spectral resolution (see Fig. \ref{fig:setup}). A diffractive optical element (DOE) is positioned near the CIS, generating a diffractogram that encodes spectral information. A neural network is trained to analyze the diffractogram, enabling accurate spectral differentiation of fluorescent beads with highly-overlapping emission profiles (see bottom-right inset in Fig. \ref{fig:setup}). Temporal alignment of the EBIS and CIS data, based on time-stamping, facilitates high-speed tracking, while simultaneously resolving two types of fluorescent beads based on the small differences in their emission spectra. This approach achieves a temporal resolution of $\sim10\thinspace\mu$s, spatial resolution of $\sim3.9\thinspace\mu$m (at the diffraction-limit of the optics used), and spectral resolution capable of distinguishing fluorescent beads with emission peaks separated by only 7 nm. In previous work, we utilized diffractive optics to achieve spectral separation of fluorescence from beads, whose emission spectra are well separated, and only for static samples \cite{wang2014hyper}. While fast spectral imaging has also been demonstrated using diffractive optics combined with computational post-processing \cite{Wang2015, Wang2018MSI}, these approaches have been restricted to macro-scale imaging. Additionally, the computational methods involved were slow and required extensive calibration.\cite{majumder2024hd}

\section{Operating Principle}
Our setup is a conventional widefield fluorescence microscope (magnification of $38\times$) with two detection paths enabled by a beam-splitter (Fig. \ref{fig:setup}). In one path, an EBIS camera (Prophesee EVK3 VGA, pixel size = $15\thinspace\mu$m) records events generated by fluorescence. In the other path, a diffractive-optical element (DOE) is placed followed by a 4f image-relay system and a CIS (Thorlabs Zelux CS165MU1, pixel size = $3.45\thinspace\mu$m). The DOE was repurposed from a different computational spectral camera, and hence, not specifically optimized for this application as it was readily available to us.\cite{majumder2024hd} Other details of the microscope are include in section 1 of the supplement. The 4f-relay is used to impart the wavelength-dependent complex transmittance of the DOE, $\exp(i2\pi h\times(n-1)/\lambda)$, where $h$ is the 2D geometry of the DOE and $n(\lambda)$ is the wavelength-dependent refractive index, onto the fluorscence image. A small gap between the intermediate image plane and the DOE increases the distinguishability of the spectra, since the Green's operator for free-space diffraction is also wavelength dependent\cite{majumder2024hd, Wang2015, Wang2018MSI, Wang2014a, Wang2014b}. In our experiments, this gap was determined empirically to minimize spatial blurring, while ensuring sufficient spectral distinguishability.\cite{Wang2018MSI} The resulting image, which we refer to as the diffractogram is then relayed onto the CIS via the 4f system. The magnified view of the diffractogram in top-right inset of Fig. \ref{fig:setup} indicates that the image is structured and this structure is wavelength dependent. Therefore, it is possible to train a classifer network to distinguish fluorescent beads based on their spectra, even when they are strongly overlapping (emission peaks are separated by only 7 nm and correlation of the two emission spectra, $\chi$ is 88\% in Fig. \ref{fig:setup}).    

\subsection{Classification based on emission spectra}
\begin{figure}[htb!]
    \centering
    \includegraphics[width=\linewidth]{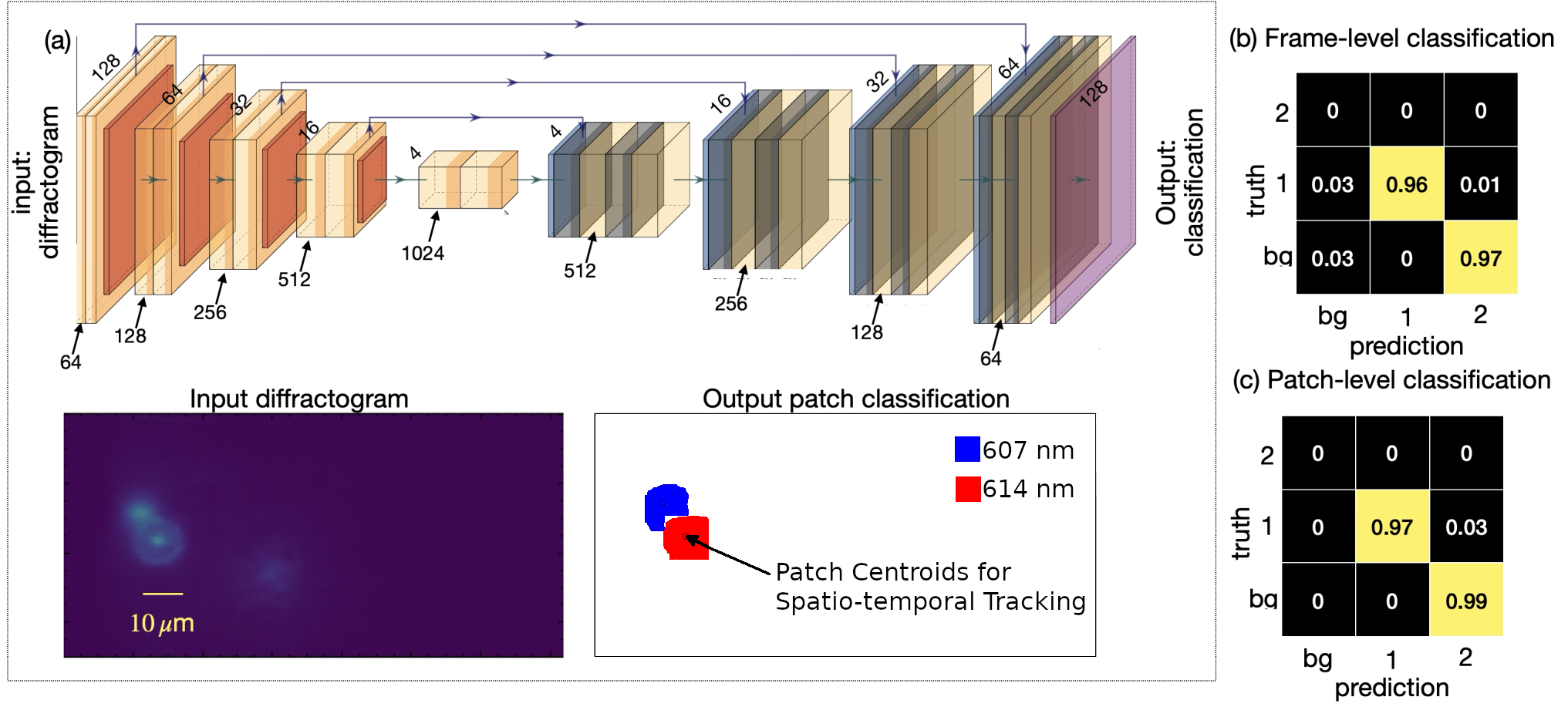}
    \caption{Classification results. (a) U-Net architecture used to analyze the diffractogram and classify beads based on their emission spectra. Bottom row shows a representative patch-level classification result for an experimental mixed-bead diffractogram. The centroid of the classified patches are used for subsequent tracking. The beads are color coded and labeled by their fluorescence peak wavelengths. (b) Confusion matrix for frame-level classification, averaged over 2,137 recorded diffractograms. The network was trained on 21,372 labeled diffractograms of pure beads. (c) Confusion matrix for patch-level classification, with a patch size of 5.8 $\mu$m $\times$ 5.8 $\mu$m, averaged over 9,000 synthetic diffractograms of mixed beads. Training was conducted on 17,000 labeled synthetic diffractograms of mixed beads.}
    \label{fig:classification}
\end{figure}

The task involves classifying beads within the field-of-view (FoV) of a microscope based on the acquired diffractogram. For this purpose, we employed a U-Net architecture with 1024 feature channels and a $3\times3$ pixel convolutional kernel \cite{ronnebergerUNetConvolutionalNetworks2015}, as illustrated in Fig. \ref{fig:classification}a. Initially, the network was trained to classify frames containing only pure beads (\emph{i.e.} beads of one emission spectra). Training data were generated by systematically translating slides containing pure beads at various concentrations using a motorized XY stage. Focus was manually adjusted as necessary throughout the data collection process (details are provided in sections 2 and 3 of the supplement). Images were captured as 16-bit raw files. Subsequently, regions without beads were excluded, and contrast in the remaining areas was enhanced before saving the processed data as 16-bit TIFF files. A corresponding matrix of class labels was stored as metadata for each image. A comprehensive description of the data collection protocol is available in the supplementary material. The network was trained with 21,372 such labeled images using a combination of cross-entropy and dice\cite{Dice2017} loss function over 5 epochs. A separate validation set of 2,137 images was used to assess performance. The trained network classified the images of pure beads with $\geq96$\% accuracy (frame-level classification) (see confusion matrix in Fig. \ref{fig:classification}b, and example images in section 4 of the supplement).

We next explored the classification of images containing a mixture of both bead types. In the absence of ground-truth data, we generated a synthetic training dataset. The diffractogram frames were first segmented into square patches of 5.8 $\mu$m $\times$ 5.8 $\mu$m (corresponding to 64 $\times$ 64 sensor pixels). Synthetic mixed-bead diffractograms were then constructed by randomly selecting these patches from experimentally recorded pure-bead diffractograms and assembling them into a single synthetic frame. The same network architecture (Fig. \ref{fig:classification}a) was trained and validated using 17,000 and 9,000 synthetic diffractograms, respectively. For patch-level classification, a polling strategy was employed to assign a class label to each 5.8 $\mu$m $\times$ 5.8 $\mu$m patch in the output. Further details of the training and validation process are provided in sections 4 and 5 of the supplement. The classification accuracy on the synthetic dataset was $\geq97$\%, averaged across both bead types (see Fig. \ref{fig:classification}c for the corresponding confusion matrix). Patch-level classification shows slightly higher accuracy than frame-level classification, likely because the former uses synthetic data, while the latter relies on experimental data.  

Finally, we applied the network, trained on synthetic data, for patch-level classification of experimentally acquired diffractograms containing mixed beads. A representative result is presented in bottom row of Fig. \ref{fig:classification}a. The input diffractogram, capturing beads with distinct emission spectra within the same FoV, is shown on the left. The network processes this input and generates a patch-classification map, identifying the bead types and their precise locations (color legends label the peak fluorescence wavelengths). The centroids of these classified patches are then extracted for subsequent spatio-temporal tracking, as detailed later. 

While frame-level classification leverages both spectral and spatial information to determine the most likely classification, patch-level classification relies solely on spectral information. This approach offers two key advantages. First, it enhances the model’s generalization capability. By excluding spatial information, the model is forced to generalize and accurately predict spectra across the entire FoV. Second, it improves computational efficiency. Processing smaller patches reduces the number of pixels to analyze, thereby decreasing the latency between input and output. In this work, we demonstrate that the model successfully generalizes to a spectral-only domain without a reduction in classification accuracy.

\begin{figure}[htb!]
    \centering
    \includegraphics[width=\linewidth]{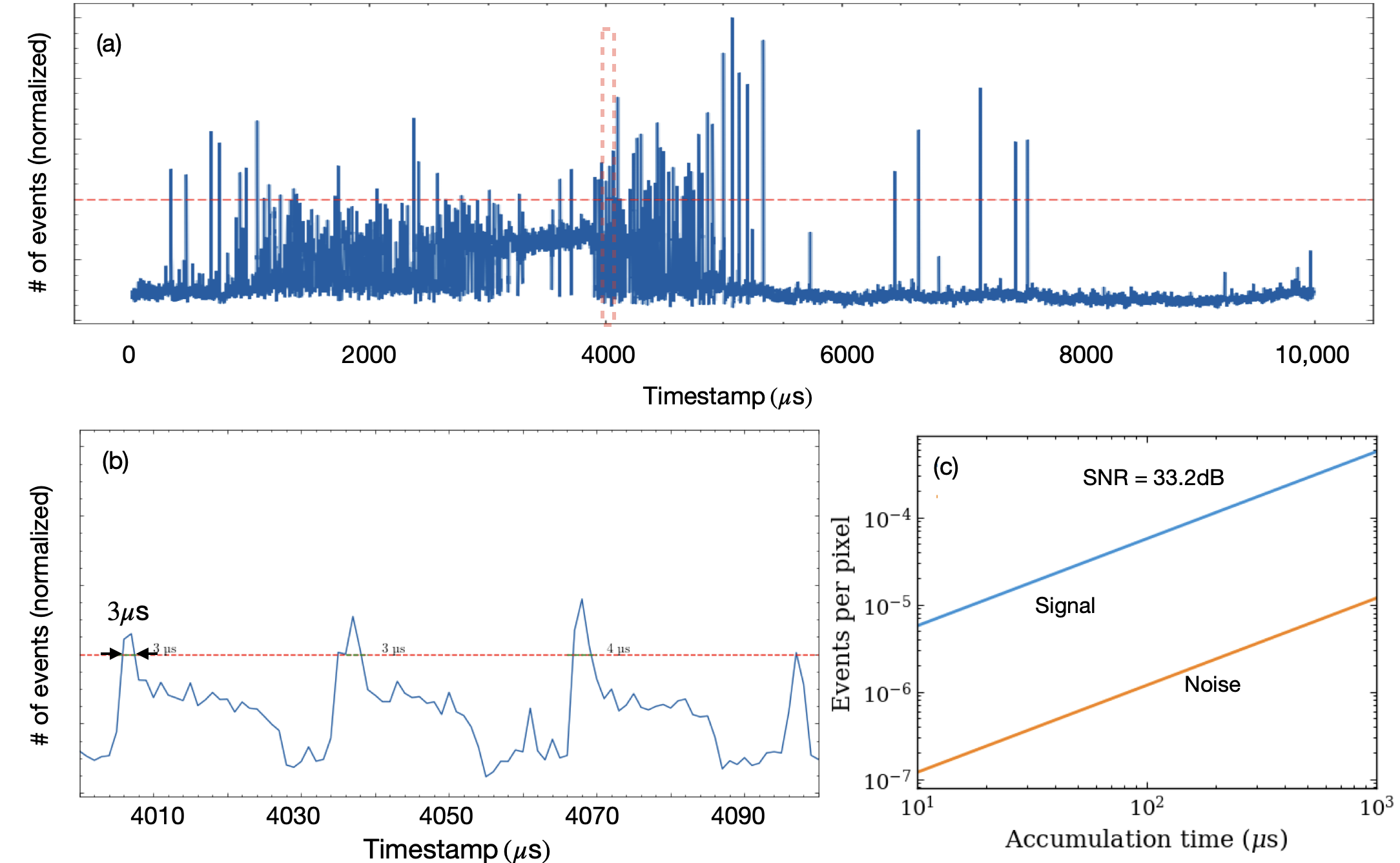}
    \caption{Estimating temporal resolution. (a) Event rate defined as total events per $\mu$s over time, sampled from the first 10 ms of recording. The orange line indicates the average event rate with no input signal.
(b) Magnified view of the orange box in (a) around $4000\thinspace\mu$s indicating three consecutive peaks and the duration of event activity above the noise floor. The average duration is 3.3 $\mu$s.
(c) Signal-to-Noise Ratio (SNR) as function of accumulation time. Signal is defined as the average events per pixel minus the baseline (no input signal), while noise is the baseline event count per pixel. The consistent SNR across accumulation times indicates that shorter accumulation periods do not compromise signal quality. The device used for the capillary flow is the CellChip\texttrademark\ (Tecan Group Ltd.). Supplementary video 1 shows the operation of the CellChip.}
    \label{fig:temporal_resolution}
\end{figure}
\subsection{Temporal resolution from event data}
Event-based sensors, also known as dynamic vision sensors (DVS), have been extensively studied for their low-latency, high-temporal resolution capabilities. Early work by Lichtsteiner \emph{et al.} \cite{lichtsteiner2008dvs} demonstrated a $128\times128$-pixel DVS with a temporal resolution of $15\thinspace\mu$s, which marked a significant advancement in asynchronous vision sensor technology. Brandli \emph{et al.} \cite{brandli2014240} further improved these sensors, achieving a latency as low as $3\thinspace\mu$s, making them suitable for high-dynamic-range applications requiring fast response times. Gallego \emph{et al.} \cite{gallego2020event} provided a comprehensive survey of event-based vision, highlighting key advances in sensor design and temporal resolution improvements, which are critical for tasks such as object tracking and neuromorphic computing. More recent efforts, such as those by Tsilikas \emph{et al.} \cite{tsilikasPhotonicNeuromorphicAccelerators2024}, have focused on leveraging photonic neuromorphic accelerators to enhance the temporal performance of these sensors for ultra-low latency applications.

Analogous to the exposure time in conventional frame-based imaging, the event-based camera employs an accumulation time, which dictates the duration over which events are aggregated before processing. Notably, this parameter is adjustable during post-processing and does not influence the actual event recording process. The EBIS camera in our setup is characterized by a nominal pixel latency of $1\thinspace\mu$s, setting the lower bound for both the accumulation time and the achievable temporal resolution. Following the approach outlined by Tsilikas \emph{et al.} \cite{tsilikasPhotonicNeuromorphicAccelerators2024}, we established a baseline noise level by operating the camera in the absence of any fluorescence signal. Upon signal introduction, the minimum accumulation time was defined as the duration above the noise threshold, determined to be $3\thinspace\mu$s (see Fig. \ref{fig:temporal_resolution}). For the majority of experiments, an accumulation time of $10\thinspace\mu$s was adopted to ensure reliable signal processing, unless otherwise noted. As indicated in Fig. \ref{fig:temporal_resolution}c, consistent SNR $>33\thinspace$dB was obtained for accumulation times as low as $1\thinspace\mu$s.

Synchronization between the CIS and the EBIS is essential to maintain temporal alignment between the captured frames and the detected events. To achieve this, the strobe output from the CIS is connected to the sync input of the EBIS, which injects precise synchronization events into the event stream at the beginning and end of each exposure. These synchronization events are subsequently used during post-processing to match the recorded events with the corresponding image frames based on their timestamps, ensuring accurate temporal correlation between the two sensor modalities.

\section{Results}
To achieve high spectral, spatial, and temporal resolution, data from the CIS and EBIS must be fused into a single, temporally and spatially aligned dataset. Ensuring that the images used for classification accurately correspond to the event data is critical. Spatial alignment was performed using a USAF focus chart, with the fields of view for both sensors calculated in micrometers and any offsets recorded prior to processing. Temporal alignment followed the method described earlier.
\begin{figure}[htb!]
    \centering
    \includegraphics[width=\linewidth]{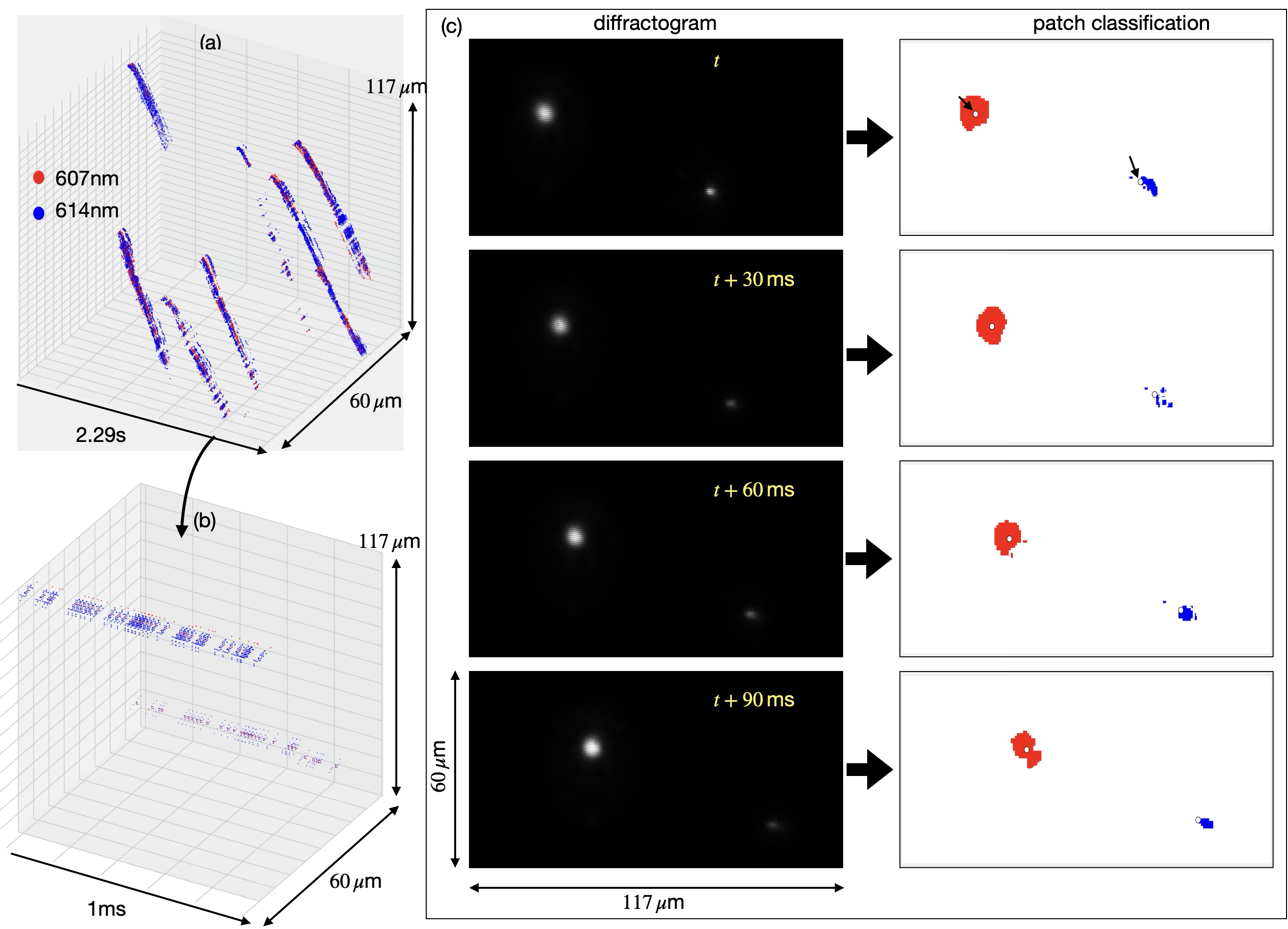}
    \caption{Spatio-temporal tracking of two-color fluorescent beads in capillary flow using combined CIS and EBIS data. (a) Trajectories of multiple beads over 2.29 s, with beads emitting at 607 nm and 614 nm shown in blue and red, respectively. (b) Magnified view of the tracking data from (a) over a 1 ms interval starting at $\sim$2.06 s, revealing finer temporal dynamics. (c) Bead classification is performed via a neural network trained on diffractograms obtained from the CIS, resulting in patch-level identification using the same color scheme as in (a). Diffractograms, captured at 30 ms intervals, show bead motion from top-left to bottom-right. Arrows on the top-right result indicate the patch centroids. Supplementary Video 2 presents the data at an effective frame rate of 10,000 frames/s.}
    \label{fig:final_results}
\end{figure}

For event-data correlation, each CIS image was thresholded, and the centroids of all classified patches in the FoV were computed (see earlier discussion). The (x, y, t) coordinates of each centroid, along with its source image, were stored in a three-dimensional spatial querying structure. Simultaneously, optical-flow analysis of the event data provided the central (x, y) coordinates and timestamps for detected objects. During each accumulation period, the (x, y, t) coordinates from the event stream were queried against the spatial structure, and the nearest neighbor was returned. This neighbor was verified to ensure the temporal difference was within 10 ms and the spatial offset within 10 $\mu$m. A Simple Online and Real-Time Tracking (SORT) algorithm \cite{Bewley2016_sort} was used to associate events to objects at the temporal resolution of the event processing. Details of this approach are provided in section 6 of the supplement.

The error margins for temporal and spatial alignment were derived from the system’s intrinsic characteristics. The 10 ms exposure time of each CIS frame introduced a potential temporal offset of up to 10 ms between events detected by the EBIS and the corresponding image frames. In terms of spatial alignment, the Euclidean distance between identical elements on the USAF chart, as observed by both the CIS and EBIS, was approximately $10\thinspace\mu$m. Once the corresponding frames were identified, they were fed into a machine learning network for classification, with the results visualized as a color-coded scatter plot along the x, y, and t axes. Figures \ref{fig:final_results}(a) and (b) summarize the results of a capillary flow experiment for time intervals of 2.29 s and 1 ms, respectively. The scatter plots clearly reveal clusters of beads, differentiated by fluorescence (peaks separated by $\sim7$ nm), moving rapidly across the field of view. From these measurements, the mean and maximum flow speeds were estimated as 9.5 mm/s and 77 mm/s, respectively. Figure \ref{fig:final_results}(c) illustrates four diffractograms, along with their corresponding patch classification maps captured at the frame rate of the CIS (30 frames/s). The centroids of these patches were extracted and tracked using the method described above. Due to the difference in frame rates between the two sensors, some events lacked corresponding CIS images. In such cases, velocity data from the optical flow algorithm was used to interpolate the classification for the missing events.

\section{Conclusion}
We have demonstrated an innovative fluorescence microscopy platform that combines event-based and conventional image sensors to achieve simultaneous high temporal, spatial, and spectral resolution. The integration of a diffractive optical element with a neural network-based classification approach enables the discrimination of closely overlapping fluorescence signals. Our system’s ability to image fast-moving fluorescent beads with sub-millisecond temporal precision highlights its potential for studying dynamic biological processes. Future work will focus on extending the system to in vivo imaging and exploring more complex biological environments.

\begin{backmatter}
\bmsection{Funding}
Chan Zuckerberg Initiative (CZI) grant: Dynamic-0000000282. 

\bmsection{Acknowledgments}
The authors thank Noor Syed for use of the EBIS camera, Richard Cavicke for the CellChip\texttrademark\ devices, and Lumos Imaging for the DOE. The support and resources from the Center for High Performance Computing at the University of Utah are gratefully acknowledged. Discussions with Al Ingold, Dajun Lin, Fernando Guevara-Vasquez and Fernando del Cueto are gratefully acknowledged.

\bmsection{Disclosures}
\noindent RM: Oblate Optics (I,E,P), Lumos Imaging (I, P).

\bmsection{Data availability} Data and code underlying the results presented in this paper are available in https://github.com/Menonlab-Rich/hyperscope

\bmsection{Supplemental document}
See Supplement 1 for supporting content. 

\end{backmatter}

\bibliography{sample}






\end{document}